# Mapping Neural Theories of Consciousness onto the Common Model of Cognition

Paul S. Rosenbloom[1], John E. Laird[2], Christian Lebiere[3], and Andrea Stocco[4]

[1]Institute for Creative Technologies & Thomas Lord Department of Computer Science,
University of Southern California, Los Angeles, CA, USA
[2]Center for Integrated Cognition, Ann Arbor, MI, USA
[3]Department of Psychology, Carnegie Mellon University, Pittsburgh, PA, USA
[4]Department of Psychology, University of Washington, Seattle, WA, USA

**Abstract.** A beginning is made at mapping four neural theories of consciousness onto the Common Model of Cognition. This highlights how the four jointly depend on recurrent local modules plus a cognitive cycle operating on a global working memory with complex states, and reveals how an existing integrative view of consciousness from a neural perspective aligns with the Common Model.

## 1    Introduction

Consciousness is a complex topic, likely due to its embedding in common-sense psychology long before it became a serious object of scientific investigation and that even once it was studied the phenomena remained poorly understood, little agreed upon, and lacking much in the way of measurable data. A recent paper [1] provides an integrative view of five neural theories that provide "prominent and complementary perspectives" on consciousness: Global Neuronal Workspace Theory [2-3]; Integrated Information Theory [4]; Recurrent Processing Theory [5]; Predictive Processing and Neurorepresentationalism [6-7]; and Dendritic Integration Theory [8]. There are many other theories [9-10], but these five are at the forefront of consciousness research in the neurosciences because of their theoretical claims – which formalize different intuitions about the nature of consciousness – and their explicit connection to brain computations.

The goal here is not to critique, choose among, or implement these theories but to begin to explore what can be learned by mapping them onto the Common Model of Cognition (CMC), a consensus abstraction of the basic building blocks underpinning human intelligence and similar forms of non-human intelligence [11]. These are neural rather than cognitive theories and so it might seem that the CMC would have little to say about them. However, as has been shown for example by work evaluating the CMC as a high-level brain architecture [12], this need not be so. Still, it is true for Dendritic Integration Theory, which focuses on activity in the dendrites of pyramidal neurons, and so nothing further will be said about it. Following a brief introduction to the CMC, each of the four remaining theories is considered in terms of how it maps onto the CMC and whether it concerns access consciousness (AC) – i.e., the ability to report – versus phenomenal consciousness (PC); that is, qualia [13].

## 2      The Common Model of Cognition

The Common Model of Cognition (CMC) [11] is an evolving community consensus concerning what must be in a cognitive architecture [14] for humanlike cognition; that is, for an architecture that either models human cognition or yields a form of cognition that is similar enough to it to be modellable in a comparable manner at a suitable level of abstraction. As implied by this, the CMC is both abstract, with no requirement for implementability, and incomplete – only including those aspects for which there is a consensus concerning necessity. Sufficiency only plays a role in helping to determine which features are considered for inclusion.

Fig. 1 shows the structure of the current consensus. It includes a working memory (WM) that acts as a hub for most of the interactions among the other modules, plus two long-term memories – a procedural memory (PM) for skills and a declarative memory (DM) for knowledge. Both long-term memories are based on symbolic data structures with associated quantitative metadata. PM can examine and modify all of WM whereas DM is limited to examining and modifying a dedicated segment in WM – i.e., the DM buffer – for retrieval cues and retrieved facts. Learning mechanisms are included for each memory – reinforcement learning (RL) [15] and procedural composition (shown as PC in the figure, but this is not phenomenal consciousness) [16-17] for PM; and structure acquisition (SA) and metadata tuning (MT) for DM. Modules also exist for perception and motor control that interact with the environment, WM (through their own buffers), and each other.

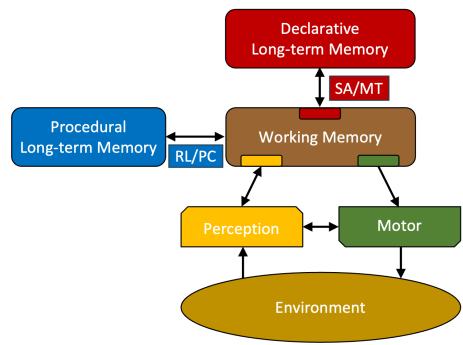

**Fig. 1.** The structure of the Common Model of Cognition.

In addition to this structural view, the CMC includes sixteen assumptions concerning how various parts of the model work. Examples include that processing is parallel both within and across modules, but that behavior is fundamentally sequential, driven by a cognitive cycle that makes decisions at ~50 ms/cycle in humans; and that complex behavior – such as planning – results from a sequence of such cycles rather than from separate modules. Additional proposals also exist for extensions to the CMC – such as for emotion [18] and metacognition [19] – but so far, a consensus is lacking on such topics.

## 3     Global Neuronal Workspace Theory

Global Workspace Theory (GWT) [2] focuses on AC via a large set of parallel modules, a global workspace (GW) – analogized to a theater stage – that broadcasts information among these modules, and a competition among the modules as to what should end up in the GW. Being in the GW is assumed necessary for information to be conscious but not necessarily sufficient. A spotlight driven by an executive function – analogized to a theater director – selects those parts that are actually to be conscious. Global Neuronal Workspace Theory (GNWT) [3] is an extension to GWT that further grounds it in the workings of the brain, including introducing a recurrent ignition process that drives the competition for presence in the GW. In this extension, the global workspace is typically identified with the prefrontal cortex, a region of the brain that fulfills the necessary requirements [20].

The natural mapping of the GW to the CMC focuses on WM, which provides the primary venue for communicating information among modules. Indeed, previous mappings of the CMC onto brain regions have already provided evidence that the WM module could be identified with the prefrontal cortex [12]. The contents of WM may thus be accessible for reporting via the joint activity of the PM, DM and motor modules. With respect to ignition, each buffered module embodies a competition about what is to go into its buffer. The role of recurrence in ignition is a further issue here, but it is addressed in Section 5, on Recurrent Processing Theory. As to the direction of the spotlight, the CMC's cognitive cycle is the basis for more integrative decisions about changes to WM, driven by PM's access to all of WM, and is thus its natural locus.

## 4     Integrated Information Theory

Integrated Information Theory (IIT) [4] focuses on PC via a set of five axiom/postulate pairs. Essentially, they state that consciousness is: (1) based on a physical system[1] that has cause-effect power on itself; (2) composed of multiple interrelated elements; (3) based on a current state with a unique cause-effect structure; (4) a unified cause-effect structure that isn't decomposable into independent components; and (5) a maximal such structure (the shape of which yields qualia). IIT is both difficult to understand and controversial [22], making some surprising predictions that call it into serious question –

---

[1] "Mathematically, a physical system constituted of $n \geq 2$ elements can be represented as a discrete time random vector of size $n$, $\mathbf{X}_t = \{X_{1,t}, X_{2,t}, \ldots, X_{n,t}\}$ with observed state at time $t$, $\mathbf{x}_t = (x_{1,t}, x_{2,t}, \ldots, x_{n,t}) \in \Omega_X$ where $(\Omega_X, d)$ is the metric space of all possible states of $\mathbf{X}_t$." [21]

such as a form of panpsychism – but key aspects of it can be viewed in terms of the CMC.

The core of the mapping depends on considering problem spaces [23], which are at the center of two instantiations of the CMC: Soar [24] and Sigma [25]. The essence of a physical system, as found in axiom/postulate pair (1) above, maps straightforwardly onto a problem space, with both a distinct current state (3) – part of WM – and a space of all possible states, with operators that move among the states. The cause-effect power on itself arises via cognitive cycles during which operators, implemented by PM, drive state changes. The states themselves are symbol structures; that is, integrations of multiple interrelated elements (2). According to IIT, the conscious part of the current state is then its maximal unified cause-effect structure (4-5).[2] Although problem spaces, as with physical systems, form the basis for process models, this is a structural theory of consciousness that does not depend on any actual processing.

## 5    Recurrent Processing Theory

Recurrent Processing Theory (RPT) [5] suggests that consciousness is a product of feedback signals. Local feedback is to yield PC whereas global feedback is to yield AC.

Local feedback, to the extent it exists, would naturally occur within modules in the CMC. In the CMC, processing in PM moves in a forward direction implying that there should not be PC within it. There is in fact backward processing for learning, but this is distinct and does not directly interact with performance. Processing in DM is omni-directional, so feedback/recurrence and thus PC should be expected there but not AC, which the CMC already rules out. The CMC presently says nothing about the nature of the processing within the perceptual and motor modules, and so it is potentially compatible with RPT there, with the arrow from WM back to perception shown in Fig. 1 initiating top-down feedback to the module. WM allows moving information in all directions among modules – with the cognitive cycle supporting continual recurrence – and so, as with the mapping of GW onto it, WM can provide global communication and thus AC. This makes the WM module compatible with the global recurrent feedback signal in the RPT. In addition, the functional abstraction of the WM module in the CMC provides a computational bridge to relate GNWT and RPT.

## 6    Predictive Processing and Neurorepresentationalism

Predictive Processing and Neurorepresentationalism (PP/NREP) [6-7] focuses on recurrent processes that mix top-down expectations and bottom-up sensory input to learn to predict the current situation and engender a hierarchy of internal representations of it. At a high level the representation is multimodal and gives rise to PC. The theory is similar to GNWT and RPT in assuming recurrence, and to GNWT as well in including a venue for integrating across modalities (although it also postulates direct intermodal communication). It differs from both – but gets closer to IIT – in focusing on how a specific form of recurrence gives rise to PC via high-level representations rather than AC via global access.

---

[2]   One aspect of this mapping that we have not been able to figure out is whether cause-effect structures would be relationships among symbol structures within individual states versus structures related across states due to what the problem space's operators test and modify.

The mapping of PP/NREP to the CMC thus combines aspects of these earlier mappings. The high-level multimodal representations are generated in WM, with the possibility of direct intermodal communication occurring within the CMC's perception and motor modules and via the bidirectional arrow that connects them in Fig. 1. Recurrence may occur both within modules and, via WM and the cognitive cycle, across modules. The complex, possibly multimodal, representations in WM should provide AC according to GNWT and RPT and PC according to IIT and PP/NREP.

Predictive computations are the foundation of PP/NREP, but are not explicit in the CMC. Still, there are aspects of both PM and DM that are implicitly predictive, as expectations about future outcomes (i.e., probabilities of rewards and events) are learned and become encapsulated in the corresponding module's metadata. The feedback from WM to perception is also capable of supporting predictive feedback. Thus, although the CMC is generally more abstract than PP/NREP, some of its components assume predictive computations that are compatible with the PP/NREP approach.

## 7    Conclusion

The last paragraphs in the previous section provide the core of a summary of how the four neural theories of consciousness in focus here map onto the CMC. In essence, they depend on the CMC supporting: (1) local processing within modules that may be recurrent; and (2) a global WM and cognitive cycle that are based on states with complex, possibly multimodal, interacting representations. Together these provide a coherent interpretation of how AC arises through WM and PC arises through either recurrent processing or complex representations. The CMC can also accommodate the spotlight from GNWT, via its cognitive cycle, and PP/NREP's direct intermodal communication.

The integrative view in [1] includes a cellular and subcellular micro-level, a local circuits and recurrent processing meso-level, a multi-area systems macro-level, and overarching concepts of richness and complexity. The analysis here essentially lays out how all of these, except for the micro-level, might be grounded in the CMC. Although the mappings described here certainly do not capture all of the subtleties of the four theories in focus here, nor may all four theories actually prove true either in their gross structure or in their details, but the hope is that the mappings here can ultimately help pave the way towards a neurocognitive theory of consciousness.

To conclude, it is worth noting how so many parts of the neural theories of consciousness discussed here are reflected in aspects of the CMC. For example, the omnidirectional and multimodal nature of the WM module is compatible with the global workspace of the GNWT, the global feedback signal of the RPT, the highest-level representations of the PP/NREP, and – in as much as it supports problem space representations – the most integrated representations in the IIT. Similarly, local computations within the non-WM modules of the CMC are compatible with the unconscious, unimodal representations of the GNWT, the local feedback signals of the RPT, and the lower-level representations of the PP/NREP. Thus, we see the CMC as a useful level of abstraction to understand the relationship between neural theories of consciousness and to extend their application to non-human and artificial agents.

**Disclosure of Interests.** The authors have no competing interests to declare that are relevant to the content of this article.

# References

1. Storm, J. F., Klink, P. C., Aru, J., Senn, W., Goebel, R., Pigorini, A., Avanzini, P., Wanduffel, W., Roelfsema, P. R., Massimini, M., Larkum, M. E., Pennartz, C. M. A. An integrative, multiscale view on neural theories of consciousness. Neuron **112**, 1531-1552 (2024)
2. Baars, B. J.: A Cognitive Theory of Consciousness. Cambridge University Press, Cambridge, UK (1988)
3. Dehaene, S., Changeux, J-P.: Experimental and Theoretical Approaches to Conscious Processing. Neuron **70**(2), 200-227 (2011)
4. Tononi, G., Boly, M., Massimini, M., Koch, C.: Integrated information theory: From consciousness to its physical substrate. Nature Reviews Neuroscience **17**, 450–461 (2016)
5. Lamme, V. A. F.: Towards a true neural stance on consciousness. Trends in Cognitive Sciences **10**(11), 494–501 (2006)
6. Hohwy, J., Seth, A.: Predictive processing as a systematic basis for identifying the neural correlates of consciousness. Philosophy and the Mind Sciences **1**(II). (2020)
7. Pennartz, C. M. A.: What is neurorepresentationalism? From neural activity and predictive processing to multi-level representations and consciousness. Behavioural Brain Research **432**, 113969 (2022)
8. Aru, J., Suzuki, M., Larkum, M. E.: Cellular mechanisms of conscious processing. Trends in Cognitive Sciences **24**(10), 814–825 (2020)
9. Seth, A. K., Bayne, T.: Theories of consciousness. Nature Reviews Neuroscience **23**, 439–452 (2022)
10. Kuhn, R. L.: A landscape of consciousness: Toward a taxonomy of explanations and implications. Progress in Biophysics and Molecular Biology **190**, 28–169 (2024)
11. Laird, J. E., Lebiere, C., Rosenbloom, P. S.: A Standard model of the mind: Toward a common computational framework across artificial intelligence, cognitive science, neuroscience, and robotics. AI Magazine **38**(4), 13-26 (2017)
12. Stocco, A., Steine-Hanson, Z., Koh, N., Laird, J. E., Lebiere, C., Rosenbloom, P. S.: Analysis of the human connectome data supports the notion of a "Common Model of Cognition" for human and human-like intelligence. Neuroimage **235**, 118035 (2021)
13. Block, N.: Two neural correlates of consciousness. Trends in Cognitive Sciences **9**(2), 46–52 (2005)
14. Kotseruba, I., Tsotsos, J. K.: 40 years of cognitive architectures: Core cognitive abilities and practical applications. Artificial Intelligence Review **53**(1), 17-94 (2020)
15. Sutton, R. S., Barto, A. G.: Reinforcement Learning: An Introduction. 2nd edn. MIT Press, Cambridge, MA (2018)
16. Anderson, J. R., Bothell, D., Byrne, M. D., Douglass, S., Lebiere, C., Qin, Y.: An integrated theory of the mind. Psychological Review **111**(4), 1036-1060 (2004)

17. Rosenbloom, P. S.: A Cognitive odyssey: From the Power Law of Practice to a general learning mechanism and beyond. Tutorials in Quantitative Methods for Psychology, **2**(2), 43-51 (2006)
18. Rosenbloom, P. S., Laird, J. E., Lebiere, C., Stocco, A., Granger, R. H., Huyck, C.: A proposal for extending the Common Model of Cognition to emotion. In Proceedings of the 22nd International Conference on Cognitive Modeling (2024)
19. Laird, J. E., Lebiere, C., Rosenbloom, P. S., Stocco: A proposal to extend the Common Model of Cognition with metacognition. arXiv:2506.07807 (2025)
20. Baars, B. J., Geld, N., Kozma, R.: Global workspace theory (GWT) and prefrontal cortex: Recent developments. Frontiers in Psychology **12**(749868) (2021)
21. Scholarpedia on Integrated Information Theory, http://www.scholarpedia.org/article/Integrated_information_theory, last accessed 2025/4/27
22. Gomez-Marin, A., Seth, A. K.: A science of consciousness beyond pseudo-science and pseudo-consciousness. Nature Neuroscience **28**(4) (2025)
23. Newell, A., Yost, G. R., Laird, J. E., Rosenbloom, P. S., Altmann, E.: Formulating the problem-space computational model. In: Rashid, R. F. (ed.) CMU Computer Science: A 25th Anniversary Commemorative, pp 255-293. ACM Press/Addison-Wesley, New York, NY (1991)
24. Laird, J. E.: The Soar Cognitive Architecture. MIT Press, Cambridge, MA (2012)
25. Rosenbloom, P. S., Demski, A., Ustun, V.: The Sigma cognitive architecture and system: Towards functionally elegant grand unification. Journal of Artificial General Intelligence **7**(1), 1–103 (2016)